\documentclass{iau-JDSS}
\usepackage{graphicx}

\title[QPO-jet relation in X-ray binaries] 
{QPO-jet relation in X-ray binaries}

\author[T. M. Belloni]   
{Tomaso M. Belloni$^1$}%

\affiliation{$^1$INAF--Osservatorio Astronomico di Brera,
Via E. Bianchi 46, I-23807 Merate, Italy \break email: tomaso.belloni@brera.inaf.it\\[\affilskip]
}

\pubyear{2009}
\volume{Volume 15}  
\pagerange{???--???}
\date{?? and in revised form ??}
\jname{Highlights of Astronomy, Volume 15}
\editors{Ian F Corbett, ed.}
\begin{document}

\maketitle

\begin{abstract}

In the past years, a clear picture of the evolution of outbursts of black-hole X-ray binaries
has emerged. While the X-ray properties can be classified into our distinct states, based on 
spectral and timing properties, the observations in the radio band have shown strong links
between accretion and ejection properties. Here I briefly outline the association between X-ray
timing and jet properties.

\end{abstract}

\firstsection 
\section{Fast time variability}

The fast time variability observed in the X-ray emission from Black-Hole Binaries (BHB)
can be extremely strong and complex. It is clearly connected to the spectral evolution
throughout their outbursts, which can be described through the use of Hardness-Intensity
Diagrams (HID; see \cite[Belloni 2009]{belloni09} and references therein).
A total fractional ms variability of $\sim$40\% is a major ``disturbance'' of the accretion 
flow that can hardly be ignored when trying to understand its properties.
Concentrating on the most basic properties, we can identify two categories: {\it loud} states
(LHS and HIMS in \cite{belloni09}), characterized by strong flat-top noise components in the
power spectra, with total fractional variability 10-40\%, and {\it quiet} states (SIMS and HSS), with less variability in the form of a power law component. Quasi-Periodic Oscillations (QPO) are observed in all
states, with a complex phenomenology. However, the HIMS-SIMS transition is very abrupt and
involves the interplay between two very different ``flavors'' of QPO. This transition can be marked in a HID
with a {it QPO line} (see Fig. \ref{fig:figure1}). At the same time, the high-energy part of the X-ray spectrum undergoes abrupt changes through the transition (see \cite[Motta et al. 2009]{motta09}).

\section{Jet ejection}

\begin{figure}
\begin{tabular}{cc}
{\includegraphics[width=6.5cm]{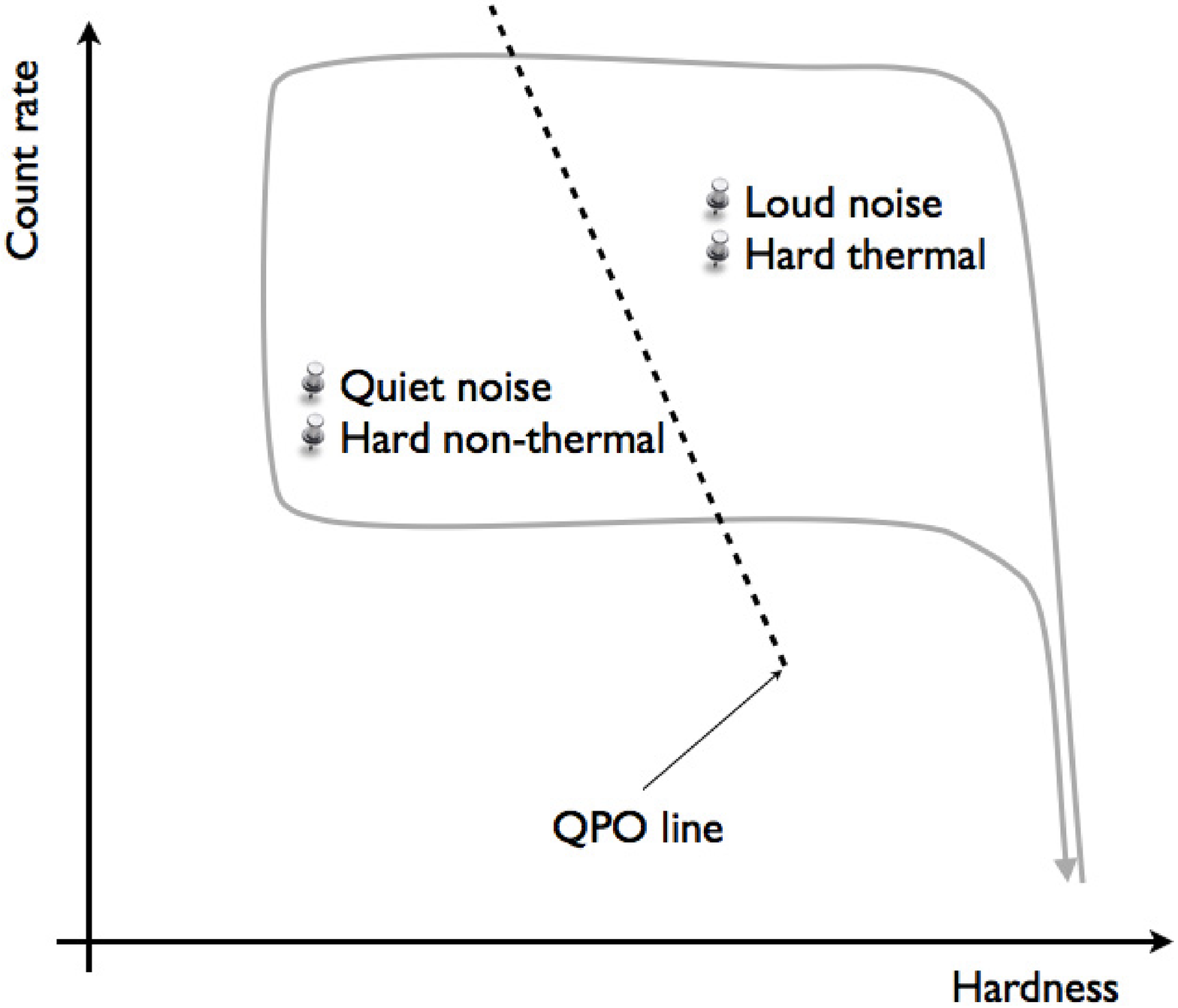}}&{\includegraphics[width=6.5cm]{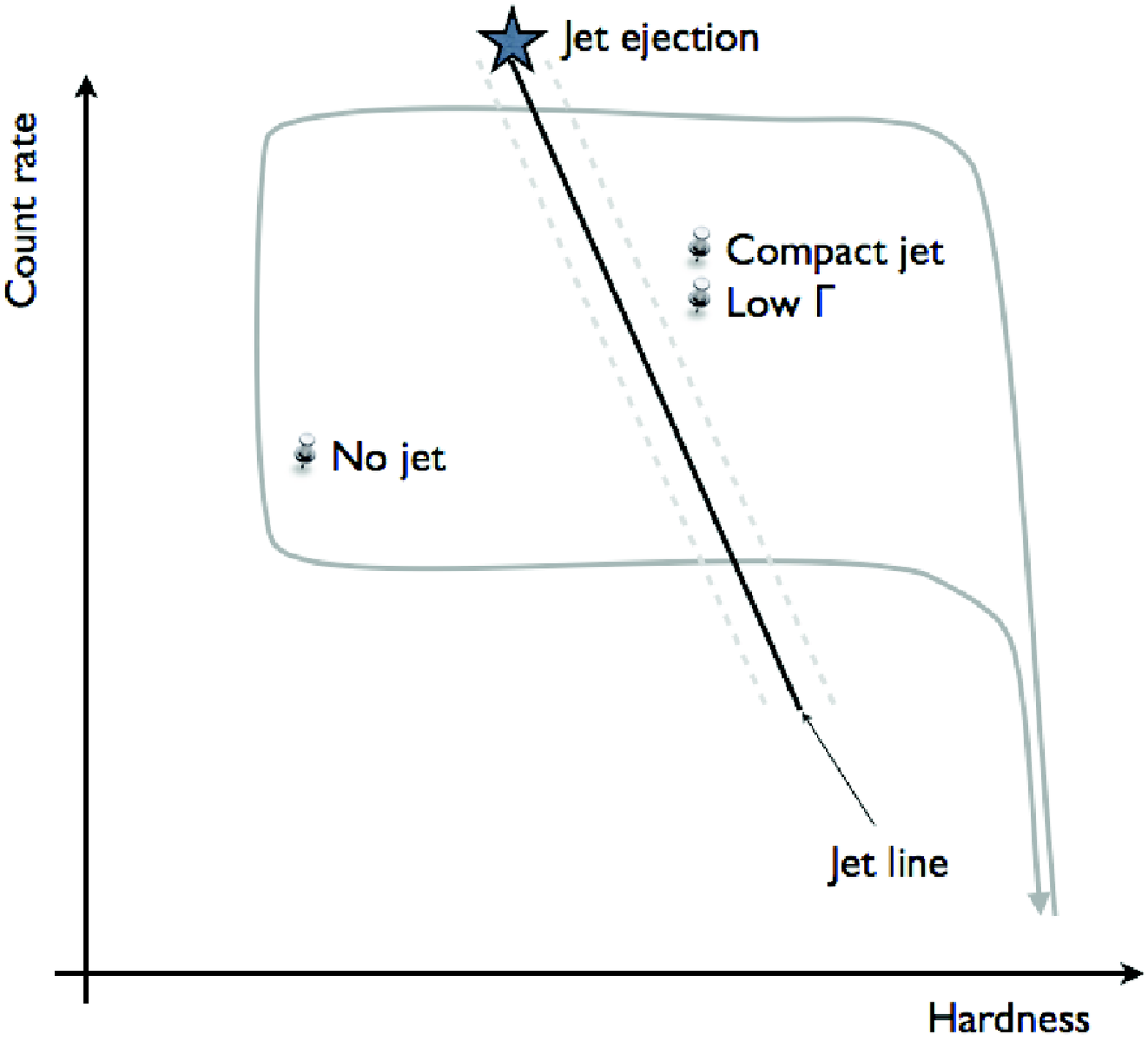}}\\
\end{tabular}
  \caption{Left: schematic HID with the two regions identified through time variability and the 
  `QPO line' between them. Right: same HID, with the radio regions and the jet line'. The two 
  fundamental lines do not coincide.}
  \label{fig:figure1}
\end{figure}

The radio properties of BHB display an evident connection with the X-ray states and transitions
(see e.g. \cite[Fender 2006]{fender06}). A relation with the states evolution was presented
by \cite[Fender, Belloni \& Gallo (2004)]{fbg04} on the basis of four well-studied systems. At its basis,
the unified picture of disk-jet coupling presented there identifies two regions of the HID: the hard 
region where a steady, compact and mildly relativistic jet is observed, and the soft region where there is no evidence of nuclear emission from the binary (see Fig. \ref{fig:figure1}). The transition between these two regions marks the ejection of a fast relativistic jet, observed as a bright radio flare or, when imaged, as a superluminal jet. The position of this transition was dubbed ``jet line''.

\section{How do they connect?}

\cite[Fender, Belloni \& Gallo (2004)]{fbg04} identified the jet line with the QPO line. This identification would lead to the attractive conclusion that the plasma responsible for the noise would be the one ejected inform of a jet. However, \cite[Fender, Homan \& Belloni (2009)]{fhb09} recently reported, on the basis of a larger sample of sources, that the two lines do not always coincide, but are close. There seems not to be a direct causal connection, as sometimes one precedes the other and vice versa. However, the close association suggests that both ejection and change in timing properties are the outcome of a complex physical transition that takes place on a longer time scale.

\section{Discussion}
The scheme outlined above, based on the HID and fast timing properties can be extended to neutron-star
systems and even to white-dwarf binaries (see \cite[Belloni 2009; K\" ording et al. 2008; Fender 2009; Tudose et al. 2009]{belloni09,kording08,fender09,tudose09}). It is clear that its properties are intimately connected to the spectral state and to the characteristics of the jet. An important key is the study of the correlated variability at optical wavelengths (see e.g. \cite{kanbach,gandhi}), which can shed light on this connection.


\begin{thebibliography}{}

\bibitem[Belloni (2009)]{belloni09}
     {Belloni, T.M.} 2009,
     in: T.M. Belloni (ed.),
     \textit{The Jet Paradigm: from Microquasars to Quasars},
     Lecture Notes in Physics (Heidelberg: Springer), in press (arXiv:0909.2474)
     
\bibitem[Fender (2006)]{fender06}
     {Fender, R.} 2006,
     in: Lewin, W.H.G. \& van der Klis, M. (eds.),
     \textit{Compact stellar X-ray sources},
     Cambridge Astrophysics Series, No. 39, Cambridge University Press, p. 381
     
     \bibitem[Fender (2009)]{fender09}
     {Fender, R.P.} 2009,
     in: T.M. Belloni (ed.),
     \textit{The Jet Paradigm: from Microquasars to Quasars},
     Lecture Notes in Physics (Heidelberg: Springer), in press (arXiv:0909.2572)
     
\bibitem[Fender, Belloni \& Gallo (2004)]{fbg04}
     {Fender, R.P., Belloni, T. \& Gallo, E.} 2004,
     \textit{MNRAS}, 355, 1105
     
      \bibitem[Kanbach et al. (2001)]{kanbach}
     {Kanbach, G., et al.,} 2004,
     \textit{Nature}, 324, 23
     
     \bibitem[Gandhi et al. (2001)]{gandhi}
     {Gandhi, P., et al.,} 2008,
     \textit{MNRAS}, 390, L29
     
     \bibitem[K\" ording et al. (2008)]{kording08}
     {K\" ording, E., et al.,} 2004,
     \textit{Science}, 320, 1318
     
\bibitem[Motta et al. (2009)]{motta09}
     {Motta, S., Belloni, T. \& Homan, J.} 2009,
     \textit{MNRAS} in press (arXiv:0908.2451)
     
     \bibitem[Tudose et al. (2009)]{tudose09}
     {Tudose, V., et al.,} 2009,
     \textit{MNRAS} in press (arXiv:0909.3604)
     
\end{thebibliography}
\end{document}